\documentclass[english,a4paper,prl,twocolumn,amsmath,amssymb,longbibliography]{revtex4-1}
\usepackage{graphicx,color,soul}

\newcommand{\dd}{\mathrm{d}}

\newcommand{\pd}[2]{\frac{\partial #1}{\partial #2}}
\newcommand{\mean}[1]{\langle #1 \rangle}
\newcommand{\Int}[1]{\int\dd #1\;}
\newcommand{\IInt}[3]{\int_{#2}^{#3}\dd #1\;}

\renewcommand{\vec}[1]{\mathbf #1}

\newcommand{\al}{\alpha}
\newcommand{\gam}{\gamma}
\newcommand{\eps}{\varepsilon}
\newcommand{\kap}{\kappa}
\newcommand{\lam}{\lambda}
\newcommand{\vhi}{\varphi}
\newcommand{\sig}{\sigma}

\newcommand{\id}{\mathbf 1}
\newcommand{\im}{\text i}
\newcommand{\x}{\vec r}

\newcommand{\Dr}{D_\text{r}}

\begin{document}

\title{From scalar to polar active matter: Connecting simulations with mean-field theory}

\author{Ashreya Jayaram}
\author{Andreas Fischer}
\author{Thomas Speck}
\affiliation{Institut f\"ur Physik, Johannes Gutenberg-Universit\"at Mainz, Staudingerweg 7-9, 55128 Mainz, Germany}

\begin{abstract}
  We study numerically the phase behavior of self-propelled elliptical particles interacting through the ``hard'' repulsive Gay-Berne potential at infinite P\'eclet number. Changing a single parameter, the aspect ratio, allows to continuously go from discoid active Brownian particles to elongated polar rods. Discoids show phase separation, which changes to a cluster state of polar domains, which then form polar bands as the aspect ratio is increased. From the simulations, we identify and extract the two effective parameters entering the mean-field description: the force imbalance coefficient and the effective coupling to the local polarization. These two coefficients are sufficient to obtain a complete and consistent picture, unifying the paradigms of scalar and polar active matter.
\end{abstract}


\maketitle


The self-organization of polar active particles into coherent dynamic patterns such as flocks and bands~\cite{vicsek12} is a robust phenomenon observed in a wide range of models~\cite{peru06,saint07,ginelli10,yang10,wens12,abkenar13,weitz15} and experiments~\cite{kudrolli08,schaller10,sumino12}. The crucial physical ingredient is the alignment of the orientations of neighboring particles (through shape, hydrodynamic coupling, etc.), a minimal model of which are self-propelled rods (SPR) either neglecting volume exclusion (point particles)~\cite{ginelli10} or assuming soft interactions that allow overlap~\cite{peru06,abkenar13,weitz15}. In contrast, active Brownian particles (ABP), respecting their finite extent but lacking alignment, aggregate into dense domains surrounded by an active gas~\cite{fily12,redn13,buttinoni13,siebert17,paliwal18,digregorio18}. The underlying mechanism has been identified as a motility-induced phase separation (MIPS) in analogy with liquid-gas coexistence, but controlled by the effective reduction of the propulsion speed in environments with an increased local density~\cite{cates15}. There has been extensive work towards a complete theoretical description of MIPS~\cite{witt14,spec15,krin19}.

While both model classes have been studied theoretically and numerically in great detail, their exact relation has received comparatively little attention. Only recently have there been several numerical studies addressing this question through varying the aspect ratio of ellipsoidal particles using different interaction models~\cite{theers18,shi18,grossmann19,vandamme19}. All studies report the suppression of MIPS and ordering into polar domains and bands as the aspect ratio is increased. However, there is still no conclusive picture on the dominant mechanism for these transitions, their relation to previously characterized phases, and how to capture them in mean-field theories.

To fill this gap, here we study a consistent model of ``dry'' anisotropic active particles. As interaction potential determining both forces and torques, we choose the repulsive Gay-Berne potential, a modified Lennard-Jones type pair potential that depends on the relative orientations of particles. This pair potential is a popular model for liquid crystals and shows a passive isotropic-nematic phase transition as the density is increased~\cite{rull95}. In addition, particles are propelled along their major axis with constant speed $v_0$~\cite{bott18}. In agreement with similar studies~\cite{shi18,grossmann19}, we find the suppression of MIPS and the emergence of polar order for elongated particles. Enhancement of MIPS in the presence of alignment is reported in Ref.~\citenum{sese18}. However, in that study an isotropic pair potential is augmented by aligning torques that do not originate from the same potential. We demonstrate that the large-scale behavior reduces to that of self-propelled rods with a density-dependent motility as introduced by Farrell \emph{et al.}~\cite{farrell12}.

\begin{figure*}[t]
  \centering
  \includegraphics{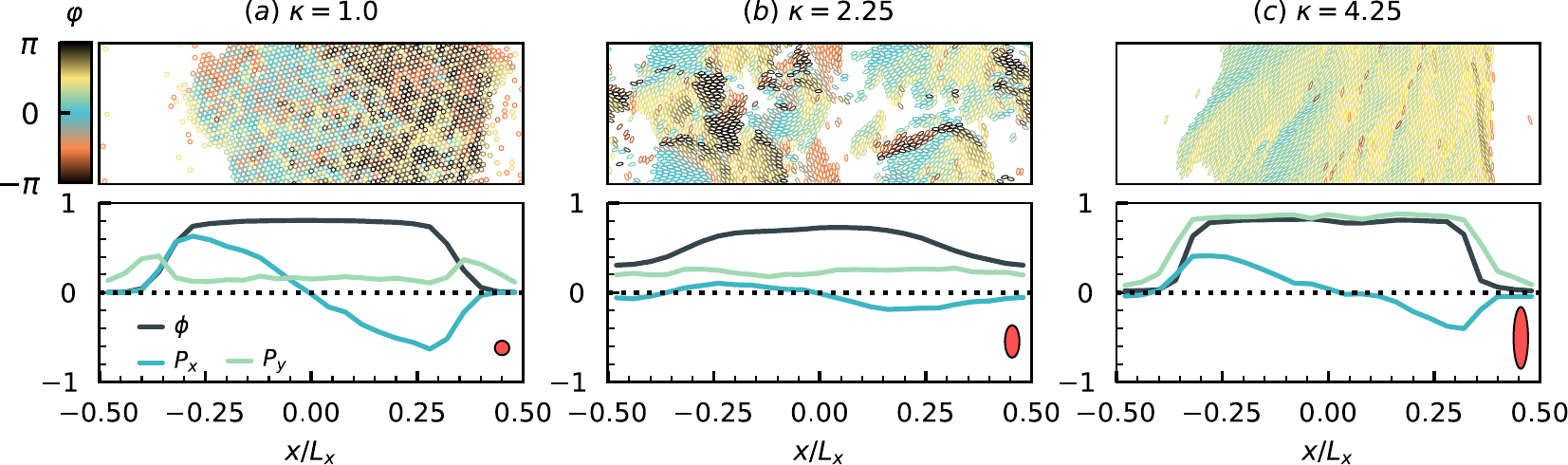}
  \caption{Simulation snapshots (top row, colors indicate the orientation of individual rods) and order parameter profiles (bottom row) at global packing fraction $\phi\simeq0.55$ for three values of the aspect ratio $\kap$. Particle shapes are indicated in the bottom right corner.}
  \label{fig:profile}
\end{figure*}


We study $N$ elongated elliptical particles with aspect ratio $\kap$ moving in two dimensions. Every particle is described by its position $\x_k$ and orientation $\vhi_k$ with unit vector $\vec e_k\equiv(\cos\vhi_k,\sin\vhi_k)^T$ corresponding to the major axis. We employ the repulsive Gay-Berne potential~\cite{gay81}
\begin{equation}
  u(\x_{kl},\vec e_k,\vec e_l) = \\ 4\eps_0\hat\eps(\xi^{-12}-\xi^{-6}), \quad \xi \equiv r/\sig_0-\hat\sig+1,
\end{equation}
where the dimensionless well-depth $\hat\eps$ and contact distance $\hat\sig$ are functions of the separation vector $\x_{kl}=\x_k-\x_l$ of two particles (with $r=|\x_{kl}|$) and their orientations. In addition, we truncate the potential at its minimum and shift it such that it smoothly approaches zero at the cut-off. Explicit expressions and further details can be found in the Supplemental Material~\cite{sm}. This interaction potential reduces to the Weeks-Chandler-Anderson potential~\cite{week72} in the limit $\kap=1$, thus making contact with previous works on motility-induced phase separation of hard discs~\cite{buttinoni13,bial15,digregorio18}.

Particles are propelled along their axis $\vec e_k$ with constant speed $v_0$. We employ an isotropic mobility $\mu_0$ with rotational mobility $3\mu_0/\sig_0^2$. In the following, we use dimensionless quantities through rescaling lengths by $\sig_0$, energy by $\eps_0$, and time by $\sig_0/v_0$. Setting the propulsion speed to $v_0=\mu_0\eps_0/\sig_0$, the equations of motion for position and orientation then read
\begin{equation}
  \label{eq:eom}
  \dot\x_k = \vec e_k - \nabla_k U, \qquad
  \dot\vhi_k = -3\pd{U}{\vhi_k}
\end{equation}
with total potential energy $U=\sum_{l<k}u(\x_{kl},\vec e_k,\vec e_l)$. We integrate Eqs.~\eqref{eq:eom} with time step $\leqslant2\times10^{-5}$ (larger $\kap$ require smaller time steps) employing periodic boundary conditions. We neglect both translational and rotational noise so that our simulations correspond to the limit of infinite P\'eclet number. The effective area covered by a single rod is $a=\pi\kap/4$, and the packing fraction is $\phi\equiv Na/(L_xL_y)$. In the following, all simulations are performed at the same global packing fraction $\phi\simeq0.55$.

Throughout, we employ a rectangular simulation box with edge lengths $L_x=3L_y$ elongated along the $x$-direction. Stable coexistence of phases separated through an interface tends to minimize the length of the interface~\cite{prest15}, even in non-equilibrium. Elongating the simulation box thus encourages the interface to align with the shorter edge, yielding systems that are translationally invariant in $y$-direction and break this symmetry in $x$-direction.

After reaching the steady state, we measure the local packing fraction $\phi(x_i)$ as well as the local polarization perpendicular
\begin{equation}
  \label{eq:Px}
  P_x(x_i) = \left\langle\frac{1}{N_i}\sum_{k\in i} \cos\vhi_k\right\rangle
\end{equation}
and parallel 
\begin{equation}
  \label{eq:Py}
  P_y(x_i) = \left\langle\left|\frac{1}{N_i}\sum_{k\in i} \sin\vhi_k\right|\right\rangle
\end{equation}
to the (average) interface. To this end, we bin particles with the central bin containing the center-of-mass of the entire system. The sums run over all $N_i$ particles in bin $i$. The angular brackets $\mean{\cdot}$ denote an ensemble average taken along trajectories of multiple independent runs. Our definition anticipates two different origins of polarity: moving aligned bands in $y$-direction corresponding to a spontaneous symmetry breaking, and jammed particles in $x$-direction (for which we retain the sign when summing).

For discs with $\kap=1$ (athermal ABP~\cite{bruss18,reich18}), Fig.~\ref{fig:profile}(a) shows a representative snapshot and the corresponding profiles of density $\phi(x)$ polarizations, $P_x(x)$ and $P_y(x)$. The state point studied here falls into the two-phase region and the system separates into a single dense domain and an active gas (although with a very low density). The interface separating both domains aligns with the shorter edge of the simulation box as expected. The density is roughly constant within both phases with a narrow interfacial region. The perpendicular polarization $P_x$ is antisymmetric, $P_x(-x)=-P_x(x)$, and peaks at the interface. The parallel polarization $P_y$ is almost constant (except for two peaks before the interface due to ``gliding'' particles) and small. It is not exactly zero because of averaging over the magnitude in Eq.~\eqref{eq:Py}.

Increasing the aspect ratio to $\kap=2.25$ [Fig.~\ref{fig:profile}(b)], we observe that the single dense domain is replaced by several transient polar domains (see also Supplemental Video 1). The average density profile is almost flat (the hump in the center is a result of the averaging with respect to the system's center of mass). The parallel polarization $P_y(x)$ is completely flat and small, indicating the absence of global polar order.
Further increasing the aspect ratio to $\kap=4.25$ [Fig.~\ref{fig:profile}(c)], the parallel polarization $P_y(x)$ now closely follows the density profile. The snapshot shows that again a single dense domain has formed but with particles that are aligned dominantly along the $y$-direction. This inhomogeneous state is stabilized by a perpendicular polarization $P_x$ that points into the dense region. Note that the migration direction of the band is parallel to the interface. This is different from the Vicsek model, where finite bands migrate perpendicular to the interface~\cite{chate08,caussin14,solon15}. The fact that these polar bands coexist with a gas has been realized by Weitz \emph{et al.}~\cite{weitz15}.

\begin{figure}[t]
  \centering
  \includegraphics{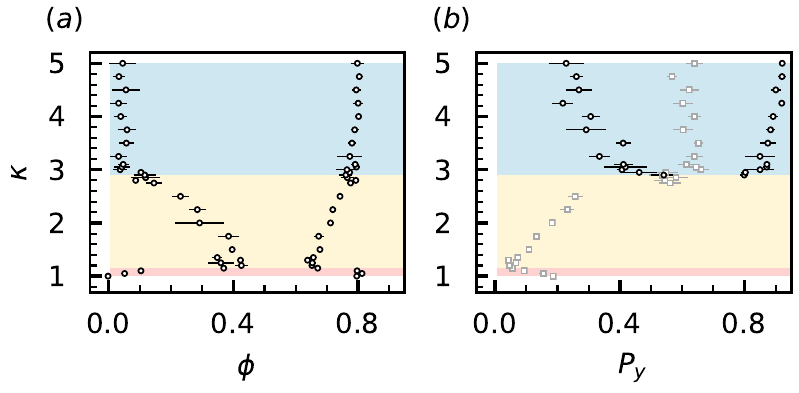}
  \caption{Coexisting (a)~packing fractions $\phi$ and (b)~parallel polarizations $P_y$ as a function of aspect ratio $\kap$ (black symbols). Also shown in (b) is the average over bins of the parallel polarization (gray symbols). The background color indicates the qualitatively distinct phases: MIPS to polar domains to polar bands (bottom to top).}
  \label{fig:sim}
\end{figure}

To summarize, we find two transitions: from coexistence without polar order (MIPS) to coexistence with polar order (polar bands) through an intermediate range of aspect ratios in which the system forms transient ``clumps'' but shows no global ordering (polar domains). In Fig.~\ref{fig:sim}, we plot the coexisting values of packing fraction $\phi$ and polarization $P_y$ extracted from fitting the profiles (see, \emph{e.g.}, Ref.~\citenum{bial15}). For small $\kap$, the MIPS coexistence region closes quickly, indicating that at $\kap\simeq1.15$ the single dense domain resolves into transient clusters. These clusters compactify and grow in size as $\kap$ is increased. The transition to polar bands at $\kap_\ast\simeq2.9$ is signaled by the sudden appearance of distinct parallel polarizations $P_y$ in the dilute and dense regions, and a jump of the average polarization. Again one single domain forms, with the coexisting densities becoming independent of the aspect ratio.

We now show how the observed collective behavior can be understood within dynamical mean-field theory~\cite{baskaran08,marc13,spec15}. To this end, we derive the coarse-grained dynamic equation for the one-body density $\psi(\x,\vhi;t)$ of a tagged particle at position $\x$ with orientation $\vhi$~\cite{sm},
\begin{equation}
  \label{eq:psi}
  \pd{\psi}{t} = -\nabla\cdot[v(\rho)\vec e\psi] - 3\pd{(\tau\psi)}{\vhi} + \Dr\pd{^2\psi}{\vhi^2}.
\end{equation}
The mean force density acting on the particle is approximated by $-\rho\zeta\vec e\psi$, where $\rho(\x,t)\equiv\IInt{\vhi}{0}{2\pi}\psi(\x,\vhi;t)$ is the local density and $\zeta$ is the force imbalance coefficient describing the reduction of the effective speed $v(\rho)\equiv 1-\zeta\rho$ due to the repulsive interactions with the surrounding particles~\cite{bial13,spec15}. For the orientations, we take into account an effective rotational diffusion $\Dr>0$ due to the coarse-graining. In the following, we rescale time $t\to t/\Dr$ and length $\x\to\x/\Dr$ to absorb $\Dr$.

\begin{figure}[t]
  \centering
  \includegraphics{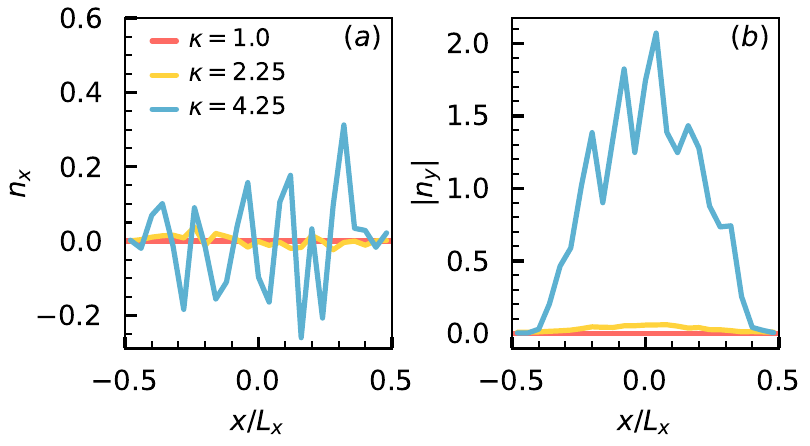}
  \caption{Components of the effective field $\vec n(x)$ determining the torque for the three aspect ratios shown in Fig.~\ref{fig:profile}.}
  \label{fig:field}
\end{figure}

In contrast to discoid particles, for $\kap>1$ there is now a mean torque density $\tau(\x,\vhi)$ changing the orientations of particles within interaction range. Its explicit expression is given in the Supplemental Material~\cite{sm}. Due to rotational symmetry in the steady state, this expression can be rearranged to $\tau=\vec n\times\vec e\equiv n_ye_x-n_xe_y$ with an effective field $\vec n(\x)$ generated by the surrounding particles. The components of this field are determined in the simulations [in analogy with Eqs.~\eqref{eq:Px} and \eqref{eq:Py}] and plotted in Fig.~\ref{fig:field} for the same aspect ratios considered in Fig.~\ref{fig:profile}. For $\kap=1$ and $\kap=2.25$, we find that both components are essentially zero. This changes for $\kap=4.25$, where $n_x$ now fluctuates more strongly around zero and $|n_y|$ (averaging the bin magnitude) develops a profile similar to the parallel polarization. Since most of the rods are aligned with the $y$-direction, it is rather demanding to gather sufficient statistics, which explains the large noise in these curves.

Nevertheless, Fig.~\ref{fig:field} provides an important insight into the structure of the mean-field equation. As argued above, polar bands correspond to a phase coexistence and the box geometry encourages an inhomogeneous density profile along the $x$-axis and a polarization along the $y$-axis. If the field depended on the density gradient, we would have $n_x\sim\partial_x\rho$. However, the observed profile does not show the expected behavior: two humps corresponding to the interface with an approximately flat value in the bulk regions. We thus conclude that the effective field $\vec n$ does not depend on the density gradient. Parallel to the interface, we find $|n_y|\sim P_y$ for polar bands.

Integrating out the orientation in Eq.~\eqref{eq:psi} yields the continuity equation $\partial_t\rho=-\nabla\cdot[v(\rho)\vec p]$ for the density with polarization density $\vec p(\x,t)\equiv\IInt{\vhi}{0}{2\pi} \vec e\psi(\x,\vhi;t)$. According to our analysis for the torque coupling, we now set $\vec n=\chi\vec p$ with some coefficient $\chi$. Note that the torque averaged over the orientations then reads $\mean{\tau}=\chi\vec p\times\vec p$, which is invariant under $\vec p\to-\vec p$. This nematic symmetry is broken by the propulsion term. The derivation of the effective hydrodynamic equations then follows Ref.~\citenum{bertin06} and yields (almost) the same evolution equation for $\vec p$ as obtained by Farrell \emph{et al.}~\cite{farrell12} for a more schematic alignment potential. This demonstrates that the exact form of the potential is not relevant and yields the same large-scale behavior. The coupling parameter entering the theory is $\gam\equiv3\chi/\Dr$.

\begin{figure}[t]
  \centering
  \includegraphics{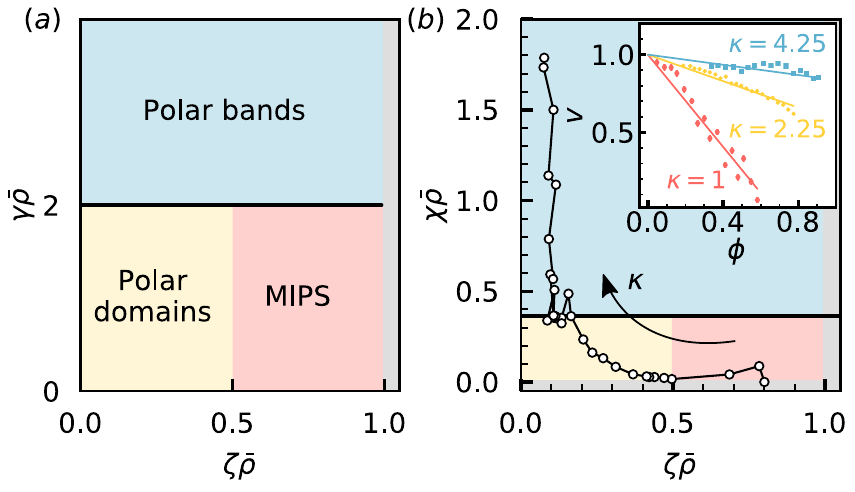}
  \caption{(a)~Mean-field phase diagram. The line between polar domains and MIPS is the spinodal from the linear stability analysis. The solid line corresponds to a critical line. (b)~Values for $\zeta$ and $\chi$ ($\propto\gam$) obtained from the simulations. The inset shows the local speed $v(\phi)$ as a function of local packing fraction for the three aspect ratios of Fig.~\ref{fig:profile}.}
  \label{fig:theory}
\end{figure}

The mean-field phase diagram is shown in Fig.~\ref{fig:theory}(a). To recapitulate, for a stationary homogeneous state with uniform density $\bar\rho=\phi/a$ and polarization $\bar p$ one finds
\begin{equation}
  \left(1-\frac{\gam\bar\rho}{2}+\frac{\gam^2\bar p^2}{8}\right)\bar p = 0,
\end{equation}
which has two solutions, $\bar p=0$ and for $\gam\bar\rho>2$ also $\bar p=(2/\gam)\sqrt{\gam\bar\rho-2}$. Linear stability analysis shows that the homogeneous polar state $\bar p>0$ is actually unstable~\cite{sm}, which agrees with the formation of polar bands in the simulations. For $\gam\bar\rho<2$, there is no global polar order. Another stability analysis of the homogeneous state with $\bar p=0$ shows that for $\zeta\bar\rho>\tfrac{1}{2}$ the system becomes spatially inhomogeneous corresponding to MIPS~\cite{cates15}, while for $\zeta\bar\rho<\tfrac{1}{2}$ it is predicted to remain homogeneous. As mentioned, in the simulations we do not find a perfectly homogeneous state but transient finite domains [cf. Fig.~\ref{fig:profile}(b)], which does agrees with the absence of a large-scale instability such as MIPS. For $\gam\bar\rho<2$, fluctuations of the polarization decay with rate $1-\gam\bar\rho/2$ due to the effective rotational diffusion. This relaxation rate is reduced as the effective torque coupling $\gam$ becomes stronger and vanishes at the transition $\gam\bar\rho=2$ to polar bands, which agrees with the transition being continuous and critical. Indeed, the larger uncertainties in Fig.~\ref{fig:sim} [and also Fig.~\ref{fig:theory}(b)] around $\kap_\ast\simeq2.9$ can be rationalized by the (in an infinite system) diverging fluctuations (see also Supplemental Video 2). In contrast, $\zeta\bar\rho=\tfrac{1}{2}$ corresponds to a spinodal (\emph{i.e.}, to the loss of linear stability).

To again make contact with the simulations, we now calculate the coefficients $\chi(\kap)$ and $\zeta(\kap)$, which are plotted in Fig.~\ref{fig:theory}(b). We estimate $\chi=|n_y|/(\rho P_y)$ from the bulk values in the dense phase for $P_y$ (Fig.~\ref{fig:profile}) and $|n_y|$ (Fig.~\ref{fig:field}). For the determination of $\zeta$, we plot the average of the projection of the velocity onto the orientation vector $v(x_i)=\mean{\sum_{k\in i}\dot\x_k\cdot\vec e_k}$ in each bin as a function of the local packing fraction, cf. inset of Fig.~\ref{fig:theory}(b). We find a linear decay, which we fit to extract the slope $\zeta$.

Starting from discs ($\kap=1$), we observe the following: for small $\kap$, the force imbalance $\zeta$ becomes smaller due to local alignment of particles, which can thus slide more easily past each other. This reduces the trapping effect responsible for the phase-separated state. At the same time, the orientional coupling $\chi$ becomes \emph{smaller}. For $\zeta\bar\rho<\tfrac{1}{2}$, the single dense domain resolves into many clusters. Further increasing $\kap$, the force imbalance $\zeta$ keeps decreasing but now the orientational coupling $\chi$ starts to grow until another phase transition takes place in which clusters merge into a single dense domain which is now highly polarized and migrates parallel to the interface. The force imbalance $\zeta\bar\rho$ in this phase is small but non-zero and independent of the aspect ratio, while $\chi$ keeps increasing.


To conclude, we have studied the phase behavior of hard and polar active particles as their aspect ratio $\kap$ is increased. We have chosen a popular pair potential modeling nematic interactions of molecular rods that reduces to the Weeks-Chandler-Andersen potential in the limit of discoid particles ($\kap=1$). We have thus made contact with the large body of theoretical studies investigating motility-induced phase separation in discoids.

In contrast to the scenario put forward in Ref.~\citenum{grossmann19}, we find numerical evidence that the effective torque couples to the polarization but not to density gradients (cf. Fig.~\ref{fig:field}). In agreement with their numerical observations, we find that motility-induced phase separation vanishes as the aspect ratio is increased, which, however, is due to a reduction of the force imbalance coefficient $\zeta(\kap)$ as particles start to align locally, and not related to torques ($\chi\simeq0$ in this regime).

Our results enforce the notion that the collective behavior of hard self-propelled rods at high P\'eclet numbers is dominated by the physics of coexistence. It can be captured by two coefficients, the force imbalance coefficient $\zeta$ and an orientational coupling $\chi$. Although the microscopic details of our system are quite different, its large-scale evolution follows the mean-field theory developed in Ref.~\citenum{farrell12} for a more schematic interaction model. Reducing the P\'eclet number (at sufficiently large aspect ratios) will eventually lead to the noise-driven transitions reported previously~\cite{ginelli10}.

Notable is the absence of vortices and chaotic (turbulent) behavior, which has been found in simulations of dry self-propelled rods~\cite{wens12} with aspect ratios comparable to those studied here. Apart from the different interaction model, an anisotropic mobility tensor is employed in that work. The importance of choosing different parallel and perpendicular mobilities has also been found in a more recent numerical work~\cite{shi18}. In Ref.~\citenum{reinken18}, to achieve chaotic behavior in a mean-field model, an additional coupling to the flow had to be assumed. It will be interesting to investigate why a reduced perpendicular mobility of rods leads to such a coupling inducing chaotic behavior.


\begin{acknowledgments}
  We acknowledge funding by the Deutsche Forschungsgemeinschaft (DFG) within collaborative research center TRR 146 (Grant No. 404840447) and the priority program SPP 1726 (Grant No. 254473714). Computations have been performed on the supercomputer MOGON II.
\end{acknowledgments}

%

\newpage

\begin{widetext}
  \section*{Supplemental Information}
\end{widetext}

\section{Repulsive Gay-Berne potential}

To model excluded volume interactions between elliptical particles, we employ the Gay-Berne potential given by~\cite{gay81}
\begin{multline}
  u_\text{GB}(\x,\vec e_k,\vec e_l) \\ = 4\eps_0\hat\eps\left[\left(\frac{1}{r/\sig_0-\hat\sig+1}\right)^{12}-\left(\frac{1}{r/\sig_0-\hat\sig+1}\right)^6\right],
\end{multline}
where $\x=\x_k-\x_l$ is the separation vector between particles $k$ and $l$ with unit orientations $\vec e_k$ and $\vec e_l$, respectively, and $r=|\x|$. The orientation-dependent well depth $\hat\eps(\hat{\x},\vec e_k,\vec e_l)$ and the orientation-dependent contact distance $\hat\sig(\hat{\x},\vec e_k,\vec e_l)$ are given by ($\hat\x=\x/r$)
\begin{equation}
  \hat\eps(\hat{\x},\vec e_k,\vec e_l) = \eps_1^\nu(\vec e_k,\vec e_l)\eps_2^\mu(\hat{\x},\vec e_k,\vec e_l)
\end{equation}
and
\begin{multline}
  \hat\sig(\hat{\x},\vec e_k,\vec e_k) = \\
  \left\{1-\frac{\al}{2}\left[\frac{\left(\hat{\x}\cdot\vec e_k + \hat{\x}\cdot\vec e_l\right)^2}{1+\al \vec e_k\cdot\vec e_l} + \frac{\left(\hat{\x}\cdot\vec e_l - \hat{\x}\cdot\vec e_k\right)^2}{1-\al \vec e_k\cdot\vec e_l}\right]\right\}^{-\frac{1}{2}},
\end{multline}
where we define
\begin{equation}
  \eps_1(\vec e_k,\vec e_l) = \sqrt{\frac{1}{1-\left(\al\vec e_k\cdot\vec e_l\right)^2}}
\end{equation}
and
\begin{multline}
  \eps_2(\hat{\x},\vec e_k,\vec e_l) = \\ 1-\frac{\al'}{2}\left\{\frac{\left(\hat{\x}\cdot\vec e_k + \hat{\x}\cdot\vec e_l\right)^2}{1+\al' \vec e_k\cdot\vec e_l} + \frac{\left(\hat{\x}\cdot\vec e_l - \hat{\x}\cdot\vec e_k\right)^2}{1-\al' \vec e_k\cdot\vec e_l}\right\}.
\end{multline}
The shape anisotropy is quantified by $\al$ defined as
\begin{equation}
  \al = \frac{\kappa^2+1}{\kappa^2-1},
\end{equation}
where $\kappa$ is the aspect ratio of the ellipse. $\al'$, a measure of anisotropy in energy, is defined as
\begin{equation}
  \al' = \frac{\kappa'^{\frac{1}{\mu}}-1}{\kappa'^{\frac{1}{\mu}}+1},
\end{equation}
where $\kappa'$ is the ratio of well depths of side-to-side and end-to-end configurations of the particle pair.

The Gay-Berne potential is thus parametrized by $\sig_0,\eps_0,\kappa,\kappa',\mu$ and $\nu$. $\sig_0$ and $\eps_0$ set the length and energy scale, respectively. We vary $\kappa$ and fix $\kappa'=5$, $\mu=1$, and $\nu=2$. Moreover, we truncate the potential at its minimum and shift it such that it smoothly approaches zero at the cut-off, \emph{i.e.},
\begin{equation}
u(\x,\vec e_k,\vec e_l) = 
\begin{cases}
  \label{eq:gb}
	u_\text{GB}(\x,\vec e_k,\vec e_l)+\eps_{\text{min}} & \text{if } r\leq r_\text{cut}\\
	0 & \text{otherwise}
\end{cases}
\end{equation}
where $\eps_\text{min}=\eps_0\hat\eps(\vec e_k,\vec e_l,r_\text{cut})$ and $r_\text{cut}/\sig_0=(2^{1/6}-1)+\hat\sig$. Note that the potential reduces to the Weeks-Chandler-Andersen potential for $\kappa=1$ (discs).

\section{Derivation of dynamic mean-field equation}

We follow Ref.~\citenum{bial13}. Going from Eqs.~(2) in the main text to the many-body evolution equation and integrating out all particles except for one, we obtain
\begin{equation}
  \partial_t\psi = -\nabla\cdot(\vec e\psi-\vec F) - 3\pd{(\tau\psi)}{\vhi}
\end{equation}
for the one-body density $\psi(\x,\vhi;t)$. The average force density on a tagged particle at position $\x$ with orientation $\vhi$ due to the surrounding particles reads
\begin{multline}
	\vec F(\x,\vhi;t) \\ = \Int{\x'\dd\vhi'} [-\nabla u(\x-\x',\vhi,\vhi')]\psi_2(\x,\vhi,\x'\vhi';t)
\end{multline}
with $\psi_2(\x,\vhi,\x'\vhi';t)$ the two-body density for two particles and $u$ the repulsive Gay-Berne pair potential Eq.~\eqref{eq:gb}. The two-body density can be represented as
\begin{equation}
  \psi_2(\x,\vhi,\x'\vhi';t) = \psi_1(\x',\vhi'|\x,\vhi;t)\psi(\x,t),
\end{equation}
where $\psi_1$ is the conditional density to find another particle at $\x'$ given a particle at position $\x$. We further factorize
\begin{equation}
  \psi_1(\x',\vhi'|\x,\vhi;t) = \rho(\x,t)g(r,\theta,\vhi'|\x)  
\end{equation}
with pair distribution function $g$ assuming that the density only changes on scales much larger than the interaction range $\sig_0$. We use a local frame in which $r$ is the distance to the other particle and $\theta$ is the angle enclosed by $\vec r-\vec r'$ with the orientation of the tagged particle.

The next step is to replace the mean force density by its projection $\vec e\cdot\vec F$ onto the orientation. We have
\begin{equation}
  \vec e\cdot\nabla u = \pd{u}{r}\cos\theta + \frac{1}{r}\pd{u}{\theta}\sin\theta.
\end{equation}
Assuming that $g(r,\theta,\vhi')$ is symmetric in $\theta$ (which should hold in the absence of vortices), the second term vanishes on integration and we obtain $\vec e\cdot\vec F=-\zeta\rho\psi$, where
\begin{multline}
  \zeta = \IInt{\vhi'}{0}{2\pi}\IInt{\theta}{0}{2\pi}\IInt{r}{0}{\infty}r\left[-\pd{u}{r}(r,\theta,\vhi')\right] \\ \times \cos\theta g(r,\theta,\vhi')
\end{multline}
is the force imbalance coefficient due to the anisotropy in the distribution of surrounding particles.

The torque density reads
\begin{multline}
  \tau(\x,\vhi;t) = \\ \Int{\x'\dd\vhi'} \left[-\pd{u}{\vhi}(\x-\x',\vhi,\vhi')\right]\psi_1(\x',\vhi'|\x,\vhi;t).
\end{multline}
Using
\begin{equation}
  -\pd{u}{\vhi} = \hat n_x\pd{e_x}{\vhi} + \hat n_y\pd{e_y}{\vhi} = -\hat n_x e_y + \hat n_ye_x = \hat{\vec n}\times\vec e
\end{equation}
with $\hat n_i(r,\theta,\vhi')=-\pd{u}{e_i}$ and average
\begin{equation}
  n_i(\x,t) = \rho(\x,t)\IInt{\vhi'}{0}{2\pi}\IInt{\theta}{0}{2\pi}\IInt{r}{0}{\infty}r\hat n_i g(r,\theta,\vhi')
\end{equation}
we can rewrite the torque density $\tau(\x,\vhi;t)=\vec n(\x,t)\times\vec e$. As the final step, we add an effective diffusion term for the orientation, which arises from interactions within the coarse-graining area. Hence,
\begin{equation}
  \pd{\psi}{t} = -\nabla\cdot[v(\rho)\vec e\psi] - 3\pd{(\tau\psi)}{\vhi} + \Dr\pd{^2\psi}{\vhi^2}
\end{equation}
with $v(\rho)=1-\zeta\rho$.

\section{Hydrodynamic equations}

We close the hierarchy of moments for the nematic tensor,
\begin{gather}
  \label{eq:Q}
  \vec Q = \vec Q_\nabla + \frac{\gam}{4}\left(\vec p\vec p^T - \frac{1}{2}|\vec p|^2\id\right), \\
  \vec Q_\nabla = -\frac{1}{16}\left[(\nabla v\vec p^T)+(\nabla v\vec p^T)^T-(\nabla\cdot v\vec p)\id\right],
\end{gather}
expressing it in terms of the polarization and its derivatives.
The final evolution equation for the polarization reads
\begin{multline}
  \label{eq:p}
  \partial_t\vec p + \lam_-(\vec p\cdot\nabla)\vec p = -\nabla\left(\frac{1}{2}v\rho\right) - \vec Q\cdot\nabla v + \frac{\gam p^2}{16}\nabla v \\ + \frac{v}{16}\nabla^2(v\vec p) - \lam_+\left[(\nabla\cdot\vec p)\vec p-\frac{1}{2}\nabla p^2\right] \\ - \left(1-\frac{\gam\rho}{2}+\frac{\gam^2p^2}{8}\right)\vec p
\end{multline}
with
\begin{equation}
  \lam_\pm(\rho) \equiv \frac{4\pm1}{16}\gam v(\rho).
\end{equation}
These equations agree with Ref.~\citenum{farrell12} up to the $\nabla^2$ term and additional gradient terms arising from $\vec Q\cdot\nabla v$.

For a homogeneous state with uniform density $\bar\rho$ and polarization $\bar p$, Eq.~\eqref{eq:p} reduces to [Eq.~(6) in the main text]
\begin{equation}
  \left(1 - \frac{\gam\bar\rho}{2} + \frac{\gam^2\bar p^2}{8}\right)\bar p = 0.
\end{equation}
For $\gam\bar\rho>2$, we consider the solution $\bar p=(2/\gam)\sqrt{\gam\bar\rho-2}$. For the next step, it is convenient to write $\gam\bar\rho=2+\eps$ and $\gam\bar p=2\sqrt\eps$. 

\section{Stability analysis of homogeneous polar state}

We now consider perturbations of density $\rho=\bar\rho+\delta\rho$ and polarization $\vec p=\bar p\vec e_y+\delta\vec p$. The linearized evolution equations become
\begin{equation}
  \partial_t\delta\rho = -\bar v\nabla\cdot\delta\vec p+\zeta\bar p\vec e_y\cdot\nabla\delta\rho  
\end{equation}
and
\begin{multline}
  \partial_t\delta\vec p = -\frac{1}{2}\left(\bar v-\zeta\bar\rho+\frac{\zeta}{2\gam}\eps\right)\nabla\delta\rho + \zeta\bar{\vec Q}\cdot\nabla\delta\rho + D\nabla^2\delta\vec p \\ - \frac{\bar v\sqrt\eps}{8}\left[5(\nabla\cdot\delta\vec p)\vec e_y-3(\partial_x\delta p_y-\partial_y\delta p_x)\vec e_x-2\nabla(\vec e_y\cdot\delta\vec p)\right] \\ + \sqrt\eps\vec e_y\delta\rho - \frac{\bar v\zeta\bar\rho\sqrt\eps}{8(2+\eps)}\vec e_y\nabla^2\delta\rho - \frac{\eps}{2}\delta\vec p - \gam\bar{\vec Q}\cdot\delta\vec p
\end{multline}
with $D\equiv\bar v^2/16$, $\bar v=1-\zeta\bar\rho>0$, and
\begin{equation}
  \gam\bar{\vec Q} = \frac{\eps}{2}\left(\begin{array}{cc}
    -1 & 0 \\ 0 & 1    
  \end{array}\right).
\end{equation}

We now gather the three fields into the vector $\vec z(\x,t)\equiv(\delta\rho,\delta p_x,\delta p_y)^T$ and consider a wave-like perturbation $\vec z\sim e^{\im qx}$. We then find $\dot{\vec z}=-\vec M\cdot\vec z$ with matrix
\begin{equation}
  \vec M = \left(\begin{array}{ccc}
  0 & \im\bar v q & 0 \\
  \im c q & Dq^2 & -5\im a q \\
  -(1+\frac{\bar v\zeta\bar\rho}{8(2+\eps)}q^2)\sqrt\eps & 5\im a q & \eps + Dq^2
  \end{array}\right),
\end{equation}
$a\equiv(\bar v/8)\sqrt\eps$ and
\begin{equation}
  c \equiv \frac{1}{2}\left(\bar v-\zeta\bar\rho+\frac{3\zeta}{2\gam}\eps\right) = \frac{1}{2}-\frac{8+\eps}{8+4\eps}\zeta\bar\rho.
\end{equation}
We calculate the characteristic polynomial $P(\lam)=\lam^3+a_2\lam^2+a_1\lam+a_0=0$ with coefficients
\begin{gather}
  a_2(q) = \eps + 2Dq^2 > 0, \\
  a_1(q) = \left(\bar v c-\frac{21}{4}\eps D\right)q^2 + D^2q^4, \\
  a_0(q) = \left(\bar v c-10D\right)\eps q^2 + \bar v\left(c-\frac{5\eps}{8+4\eps}\zeta\bar\rho\right)Dq^4.
\end{gather}
The Routh-Hurwitz conditions for stability are $a_2>0$, $a_0>0$, and $a_2a_1>a_0$. Inspecting $a_0$, we see that
\begin{equation}
  \eps > \frac{6\zeta\bar\rho+2}{3\zeta\bar\rho-1}, \qquad
  \zeta\bar\rho > \frac{1}{3}
\end{equation}
for $a_0>0$ for all $q$ to hold. The condition $a_2a_1-a_0>0$ becomes
\begin{multline}
  \left(10-\frac{21}{4}\eps\right)D\eps q^2 \\ + \left[\bar v\left(\frac{1}{2}-\frac{8-4\eps}{8+4\eps}\zeta\bar\rho\right)-\frac{19}{2}\eps D\right]D q^4 \\ + 2D^3q^6 > 0.
\end{multline}
The square term is positive only for $\eps<\tfrac{40}{21}\simeq1.9$. Hence, both conditions cannot be fulfilled simultaneously for small $q$ and thus the homogeneous state is unconditionally unstable and decays into polar bands.

\end{document}